# Coherent Potential Approximation as a Voltage Probe

M. Ye. Zhuravlev,[1,2] A. V. Vedyayev,[3] K. D. Belashchenko,[1] and E. Y. Tsymbal[1*]

[1] *Department of Physics and Astronomy, Nebraska Center for Materials and Nanoscience, University of Nebraska, Lincoln, Nebraska 68588, USA*

[2] *Kurnakov Institute for General and Inorganic Chemistry, Russian Academy of Sciences, 119991 Moscow, Russia*

[3] *Department of Physics, M. V. Lomonosov Moscow State University, 119899 Moscow, Russia*

**Abstract**

Coherent potential approximation (CPA) has widely been used for studying residual resistivity of bulk alloys and electrical conductivity in inhomogeneous systems with structural disorder. Here we revisit the single-site CPA within the Landauer-Büttiker approach applied to the electronic transport in layered structures and show that this method can be interpreted in terms of the Büttiker's voltage-probe model that has been developed for treating phase breaking scattering in mesoscopic systems. We demonstrate that the on-site vertex function which appears within the single-site CPA formalism plays a role of the local chemical potential within the voltage-probe approach. This interpretation allows the determination of the chemical potential profile across a disordered conductor which is useful for analyzing results of transport calculations within the CPA. We illustrate this method by providing several examples. In particular, for layered systems with translational periodicity in the plane of the layers we introduce the local resistivity and calculate the interface resistance between disordered layers.

PACS: 73.63.-b, 73.23.-b, 73.50.Bk, 73.40.-c

# 1. Introduction

Coherent potential approximation (CPA)[1] is a powerful method for studying materials where substitutional disorder determines their electronic properties. The CPA results from the self-consistent solution of the quantum-mechanical multiple-scattering problem, which allows treating disorder in terms of the configuration-averaged scattering matrix. Typically such a solution is obtained within the single-site approximation, in which the properties of all sites but one in the system are averaged over, and that one is treated exactly.[2] This approach has been broadly used for the description of short-ranged scattering in binary alloys.[3,4] The CPA goes beyond the limits of low concentration and weak scattering in a physically realistic way, providing self-consistency of the solution.

The CPA has also become a useful tool for studying transport properties of alloys and disordered systems within the linear response theory.[5] The configurational averaging in this case requires averaging of the two one-electron Green's functions $\langle GG \rangle$, in contrast to equilibrium properties that are simply determined by $\langle G \rangle$. Thus, the extension to the linear response involves the determination of the so-called "vertex corrections" $\langle GG \rangle - \langle G \rangle \langle G \rangle$. This calculation can be performed consistently with the single-site CPA resulting in a closed set of equations for the conductance.[5] Such an approach has been extensively used for calculating the residual resistivity of binary alloys[6,7,8,9] and layered structures,[10,11] including the extension to treat realistically band structures of disordered systems.[12,13,14,15,16,17]

The configurational averaging within the CPA substitutes an effective medium that possess translational invariance for the original system which is characterized by a random non-periodic potential. The CPA replaces the latter by the self-energy $\Sigma$, which is an energy-dependent non-Hermitian operator. Its real part shifts the energy levels of the undisturbed system, whereas the imaginary part broadens the energy levels due to the finite scattering lifetime. Thus, the original system which in the absence of inelastic scattering would describe phase-coherent propagation of electronic waves is replaced by an artificial system that involves phase non-conserving scattering.

This situation is analogous to that within the Büttiker's voltage-probe model introduced to treat dephasing in mesoscopic physics.[18] This model involves fictitious voltage probes into an otherwise coherent system, which produce phase-breaking processes. No net current flows in the fictitious electrodes, and hence all electrons scattered into the voltage probes are emitted back into the sample. Such a scattering process is incoherent, and phase memory of the scattered electrons is completely lost. To realize this model in a practical calculation, fictitious probes are attached to each site of the sample, and their chemical potentials are adjusted such that no net current flows in the fictitious electrodes. Due to its appealing simplicity the voltage-probe model has been extensively used for studying quantum transport in quantum dots,[19,20] molecular junctions,[21,22] nanowires[23,24] and other mesoscopic and nano systems.[25,26,27,28] It has been shown that there is an analogy between the voltage-probe method and imaginary-potential models for dephasing.[29]

In this paper we revisit the CPA[1,5] within the Landauer-Büttiker approach[30,31] applied to the electronic transport in layered structures and show that this method can be interpreted in terms of the voltage-probe model.[18] We demonstrate that the on-site vertex function which appears within the single-site CPA formalism plays a role of a local chemical potential within the voltage-probe approach. This interpretation allows the determination of a chemical potential profile across a disordered sample which is useful for analyzing results of transport calculations within the CPA.

In particular, for layered systems with translational periodicity in the plane of the layers we demonstrate the possibility of introducing the local resistivity and calculating the interface resistance between disordered layers.

The paper is organized as follows. In section 2 we briefly outline the Büttiker's voltage-probe model. In section 3 we revisit the single-site CPA, and in section 4 derive expressions for transmission within the Landauer-Büttiker approach. In section 5 we show that the CPA results can be interpreted within the voltage-probe model. In section 6 we perform calculations for particular layered systems. In section 7 we summarize the results.

**2. Voltage-probe model**

Following the Landauer-Büttiker approach [30,31] we consider a "sample" attached to two semi-infinite electrodes. The electrodes are connected to reservoirs that are characterized by the equilibrium Fermi distribution functions $f_L(E) = f(E - \mu_L)$ and $f_R(E) = f(E - \mu_R)$, where $E$ is the energy and $\mu_L$ and $\mu_R$ are the chemical potentials of the left and right electrodes respectively. The electric current is driven in the sample by the applied voltage $V$, such that $\mu_L - \mu_R = eV$.

Within the Büttiker's voltage-probe model,[18] each atomic site of the sample is connected to a fictitious electrode $n$ that is characterized by the equilibrium Fermi function $f_n(E) = f(E - \mu_n)$ with chemical potential $\mu_n$. The electrodes are allowed to float to different $\mu_n$, to ensure zero current in the electrodes and thus local current conservation in the system.

Within the linear response the electric current at zero temperature in electrode $p$ ($p = L, R,$ or $n$) is given by (see, e.g., ref. 32)

$$I_p = \frac{2e}{h} \sum_q T_{pq} (\mu_p - \mu_q), \quad (1)$$

where summation is performed over all electrodes $q$ ($q = L, R,$ or $n$) and $T_{pq}$ is the transmission between electrodes $q$ and $p$. The latter can be calculated as follows

$$T_{pq} = -Tr\left[(\Sigma_p - \Sigma_p^\dagger)\bar{G}(\Sigma_q - \Sigma_q^\dagger)\bar{G}^\dagger\right], \quad (2)$$

where $\Sigma_p$ and $\Sigma_q$ are the self-energies associated with $p$ and $q$ electrodes respectively and $\bar{G}$ is the (retarded) Green's function[33] of the sample coupled to the electrodes

$$\bar{G}(E) = \left[E - H - \Sigma_L(E) - \Sigma_R(E) - \sum_n \Sigma_n\right]^{-1}. \quad (3)$$

Here $\Sigma_L(E)$, $\Sigma_R(E)$, and $\Sigma_n$ are the self-energies associated with the left, right and floating electrodes, and summation is performed over floating electrodes (scattering sites of the sample). The electric current (1) is obtained from transmission (2) calculated at the Fermi energy $E = E_F$.

Chemical potentials of the floating electrodes are found by assuming that the electric current flowing in each electrode is zero. Using Eq. (1) we find

$$0 = T_{nL}(\mu_n - \mu_L) + T_{nR}(\mu_n - \mu_R) + \sum_m T_{nm}(\mu_n - \mu_m). \quad (4)$$

This equation may be interpreted as a local current conservation condition. The solution of this system of linear equations determines chemical potentials at each site of the sample. Once the chemical potentials are found the net current passing through the sample is given by the current in the left (or right) electrode:

$$I = I_L = \frac{2e}{h}\left\{T_{LR}(\mu_L - \mu_R) + \sum_n T_{Ln}(\mu_L - \mu_n)\right\}. \tag{5}$$

Thus, the voltage-probe model introduces phase breaking scattering in the system ensuring current conservation throughout the sample.

Below we apply this approach to layered structures which are infinite and translationally periodic in the plane of the layers. In this case it is convenient to introduce the transverse wave vector $\mathbf{k}_\parallel = (k_x, k_y)$ that is conserved during transmission across the sample. The Green's function of the system $\bar{G}(\mathbf{k}_\parallel, E)$, as well as the self-energies of the left and right electrodes, $\Sigma_L(\mathbf{k}_\parallel, E)$ and $\Sigma_R(\mathbf{k}_\parallel, E)$, become functions of $\mathbf{k}_\parallel$. We use index $n$ to characterize the layer number, i.e. $n = 1,...N$, where $N$ is the total number of layers, and assume that the self-energies of the floating electrodes are layer dependent but constant within each layer. In this case the summation over in-plane sites can be replaced by the respective integrations, so that the transmission functions entering Eqs.(4) and (5) above are given by

$$\begin{aligned}
T_{LR} &= \int \Sigma_L^I(\mathbf{k}_\parallel) \bar{G}_{1N}(\mathbf{k}_\parallel) \Sigma_R^I(\mathbf{k}_\parallel) \bar{G}_{N1}^\dagger(\mathbf{k}_\parallel) \frac{d\mathbf{k}_\parallel}{(2\pi)^2}, \\
T_{Ln} &= \int \Sigma_L^I(\mathbf{k}_\parallel) \bar{G}_{1n}(\mathbf{k}_\parallel) \Sigma_n^I \bar{G}_{n1}^\dagger(\mathbf{k}_\parallel) \frac{d\mathbf{k}_\parallel}{(2\pi)^2}, \quad n = 1, 2,...N, \\
T_{nR} &= \int \Sigma_n^I \bar{G}_{nN}(\mathbf{k}_\parallel) \Sigma_R^I(\mathbf{k}_\parallel) \bar{G}_{Nn}^\dagger(\mathbf{k}_\parallel) \frac{d\mathbf{k}_\parallel}{(2\pi)^2}, \quad n = 1, 2,...N, \\
T_{mn} &= \int \Sigma_m^I \bar{G}_{mn}(\mathbf{k}_\parallel) \Sigma_n^I \bar{G}_{nm}^\dagger(\mathbf{k}_\parallel) \frac{d\mathbf{k}_\parallel}{(2\pi)^2}, \quad m, n = 1, 2,...N.
\end{aligned} \tag{6}$$

Here subscripts in the Green's function $\bar{G}$ denote its matrix elements between different sites, and we implicitly assume a single band model. In Eq. (6) we have defined $\Sigma_R^I = 2\,\text{Im}\,\Sigma_R$, $\Sigma_L^I = 2\,\text{Im}\,\Sigma_L$, and $\Sigma_n^I = 2\,\text{Im}\,\Sigma_n$. We have also assumed that the left and right electrodes are coupled to the sample at sites 1 and $N$, respectively, so that the respective matrix elements of the self-energy operators are $\langle n|\Sigma_L|m\rangle = \Sigma_L \delta_{n1}\delta_{m1}$ and $\langle n|\Sigma_R|m\rangle = \Sigma_R \delta_{nN}\delta_{mN}$.

## 3. Single-site CPA

Now we outline the single-site CPA. We assume that there is a random potential $U$ on each site in the sample region. The CPA replaces the disordered system by an effective medium that is described by complex coherent potential (self-energy) $\Sigma = \sum_n \Sigma_n$, the components $\Sigma_n$ being dependent on layer $n = 1...N$, but independent of a site within the layer. The self-consistency condition assumes that this Green's function of the effective medium $\bar{G}$ is equal to the Green's function $\langle G \rangle$ averaged over disorder configurations, that is

$$\bar{G} = \langle G \rangle, \tag{7}$$

where $\langle ... \rangle$ denotes averaging over disorder configurations. This provides condition to find the coherent potential $\Sigma$. Considering $(U - \Sigma)$ as a perturbation, $G$ can be written in terms of $\bar{G}$

$$G = \bar{G} + \bar{G}(U - \Sigma)G. \tag{8}$$

Averaging Eq. (8) and taking into account Eq. (7) we find

$$\langle (U - \Sigma)G \rangle = 0. \tag{9}$$

Equivalently this equation can be expressed in terms of the $T$ matrix[34] which is defined by

$$G = \bar{G} + \bar{G}T\bar{G}, \tag{10}$$

$$T = (U - \Sigma) + (U - \Sigma)\bar{G}T, \tag{11}$$

and implies that

$$\langle T \rangle = 0. \tag{12}$$

This equation can be solved within the single-site CPA, which introduces a single-site $T$-matrix $T_i$ according to

$$T_i = (U_i - \Sigma_i) + (U_i - \Sigma_i)\bar{G}T_i, \tag{13}$$

where $U_i$ and $V_i$ are on-site random and coherent potentials. As follows from the multiple-scattering theory (see, e.g., ref. 35), $T$ may be written as a sum of single-site contributions:

$$T = \sum_i Q_i, \tag{14}$$

$$Q_i = T_i \left(1 + \bar{G} \sum_{j \neq i} Q_j\right). \tag{15}$$

These equations have a simple physical interpretation. The total scattered wave is a sum of contributions from each atom given by the atomic $T$-matrix applied on an effective wave. The effective wave consists of the incident wave and of the contribution to the scattered wave from all other sites.

The single-site approximation assumes that the statistical correlation of $T_i$ and of the corresponding effective wave are negligible. Then we can decouple the averaging in Eq. (15) so that Eqs. (14) and (15) average to

$$\langle T \rangle = \sum_i \langle Q_i \rangle, \tag{16}$$

$$\langle Q_i \rangle = \langle T_i \rangle \left(1 + \bar{G} \sum_{j \neq i} \langle Q_j \rangle\right). \tag{17}$$

Thus the self-consistency condition (12) becomes

$$\langle T_i \rangle = 0. \tag{18}$$

In our case, we have $N$ non-equivalent sites within the sample, and hence Eq.(18) represents a set of $N$ coupled non-linear equations. For example, if we assume that disorder is formed by a binary alloy characterized by on-site energy $U^A$ with a probability $q^A$ and on-site energy $U^B$ with a probability $q^B$ ($q^A + q^B = 1$), Eq. (18) reads

$$q^A \frac{U^A - \Sigma_n}{1 - (U^A - \Sigma_n)\bar{G}_{nn}} + q^B \frac{U^B - \Sigma_n}{1 - (U^B - \Sigma_n)\bar{G}_{nn}} = 0, \quad n = 1, 2, \ldots N . \tag{19}$$

Here $\bar{G}_{nn}$ is the on-site Green's function

$$\bar{G}_{nn}(E) = \int \bar{G}_{nn}(\mathbf{k}_\parallel, E) \frac{d\mathbf{k}_\parallel}{(2\pi)^2} , \tag{20}$$

where

$$\bar{G}(\mathbf{k}_\parallel, E) = \left[ E - H - \Sigma_L(\mathbf{k}_\parallel, E) - \Sigma_R(\mathbf{k}_\parallel, E) - \sum_n \Sigma_n \right]^{-1} . \tag{21}$$

Eqs. (19) can be used to find $\Sigma_n$. Normally these equations are solved numerically using an iterative procedure.

## 4. CPA transmission

Transmission across a disordered sample requires averaging over disorder. Using Eq. (2) for transmission between left and right electrodes across the sample that is described by Green's function $G$ we obtain

$$T = \langle T_{LR} \rangle = Tr\left[ \Sigma_L^I \langle G \Sigma_R^I G^\dagger \rangle \right], \tag{22}$$

where we took in to account the fact that operator $\Sigma_L$ is configuration-independent. The average $\langle G \Sigma_R^I G^\dagger \rangle$ entering Eq. (22) can be calculated within the single-site CPA using the approach developed by Velický.[5] Using Eqs. (10) for the Green's function is terms of the $T$-matrix, and taking into account that $\Sigma_R$ does not depend on a random configuration and that according to Eq. (12) $\langle T \rangle = 0$, we find

$$\langle G \Sigma_R^I G^\dagger \rangle = \bar{G} \Sigma_R^I \bar{G}^\dagger + \bar{G} \Gamma \bar{G}^\dagger . \tag{23}$$

Here operator $\Gamma$ known as the vertex correction is defined by

$$\Gamma = \langle T \bar{G} \Sigma_R^I \bar{G}^\dagger T^\dagger \rangle . \tag{24}$$

Next, using Eqs.(16) and (17), $\Gamma$ can be represented as

$$\Gamma = \sum_{nm} \Gamma_{nm} , \tag{25}$$

where

$$\Gamma_{nm} = \langle Q_m \bar{G} \Sigma_R^I \bar{G}^\dagger Q_n^\dagger \rangle , \tag{26}$$

or

$$\Gamma_{nm} = \left\langle T_m \left(1 + \bar{G}\sum_{l \neq m} Q_l \right) \bar{G}\Sigma_R^I \bar{G}^\dagger \left(1 + \sum_{s \neq n} Q_s^\dagger \bar{G}^\dagger \right) T_n^\dagger \right\rangle. \tag{27}$$

Consistent with the single-site approximation of CPA we decouple Eq.(27) as follows

$$\Gamma_{nm} \approx \left\langle T_m \left\langle \left(1 + \bar{G}\sum_{l \neq m} Q_l \right) \bar{G}\Sigma_R^I \bar{G}^\dagger \left(1 + \sum_{s \neq n} Q_s^\dagger \bar{G}^\dagger \right) \right\rangle T_n^\dagger \right\rangle. \tag{28}$$

Now we can take into account Eq. (18) saying that $\langle T_i \rangle = 0$ and the fact that variations of $T_n$ on different sites are statistically independent, so that $\langle T_m ... T_n^\dagger \rangle = \delta_{mn} \langle T_n ... T_n^\dagger \rangle$. In a similar way we can conclude that $\langle Q_l \rangle = 0$ and $\langle Q_l ... Q_s^\dagger \rangle = \delta_{ls} \langle Q_l ... Q_l^\dagger \rangle$, which leads to

$$\Gamma_{mn} = \delta_{mn}\Gamma_n = \delta_{mn} \left\langle T_n \left[ \bar{G}\Sigma_R^I \bar{G}^\dagger + \sum_{l \neq n} \bar{G}\langle Q_l \bar{G}\Sigma_R^I \bar{G}^\dagger Q_l^\dagger \rangle \bar{G}^\dagger \right] T_n^\dagger \right\rangle. \tag{29}$$

Using the same approximation in Eq. (26) we obtain from Eq. (29)

$$\Gamma_n = \left\langle T_n \left[ \sum_{l,m} \bar{G}_{nl} (\Sigma_R^I)_{lm} \bar{G}_{mn}^\dagger + \sum_{l \neq n} \bar{G}\Gamma_l \bar{G}^\dagger \right] T_n^\dagger \right\rangle. \tag{30}$$

This is a system of equations which can be solved using an appropriate basis. For example, within a single-band tight-binding model we define

$$\Gamma_n = |n\rangle \gamma_n \langle n|, \tag{31}$$

and

$$T_n = |n\rangle t_n \langle n|. \tag{32}$$

Thus, Eq. (30) is reduced to

$$\gamma_n \left[ 1 + \langle t_n t_n^\dagger \rangle \bar{G}_{nn} \bar{G}_{nn}^\dagger \right] = \langle t_n t_n^\dagger \rangle \left[ \sum_{l,m} \bar{G}_{nl} (\Sigma_R^I)_{lm} \bar{G}_{mn}^\dagger + \sum_l \bar{G}_{nl} \gamma_l \bar{G}_{ln}^\dagger \right], \tag{33}$$

where we included the diagonal term in the summation. We note that here indices $l$, $m$, and $n$ refer to sites and do not take into account periodicity of our system in the plane. Once the vertex function is found the transmission can be obtained using Eqs.(22) and (23).

## 5. CPA as a voltage probe

Now, using the single-site CPA formalism we prove the following identity

$$\text{Im}\,\Sigma_n \left[ 1 + \langle t_n t_n^\dagger \rangle |\bar{G}_{nn}|^2 \right] = \langle t_n t_n^\dagger \rangle \text{Im}\,\bar{G}_{nn}. \tag{34}$$

According to Eq.(13) we have

$$(U_n - \Sigma_n) = T_n \frac{1}{1 + T_n \bar{G}_{nn}} = T_n \left(1 + T_n \bar{G}_{nn}\right)^\dagger \frac{1}{\left|1 + T_n \bar{G}_{nn}\right|^2}. \tag{35}$$

Taking an imaginary part and restructuring the terms we obtain

$$-\mathrm{Im}\,\Sigma_n\left[1+T_n\bar{G}_{nn}+\bar{G}_{nn}^{\dagger}T_n^{\dagger}+T_n\bar{G}_{nn}\bar{G}_{nn}^{\dagger}T_n^{\dagger}\right]=\mathrm{Im}\,T_n+\mathrm{Im}\left(T_n\bar{G}_{nn}^{\dagger}T_n^{\dagger}\right). \qquad (36)$$

Averaging over random configurations and taking into account $\langle T_n\rangle=0$, we find

$$\mathrm{Im}\,\Sigma_n\left[1+\left\langle T_n\left|\bar{G}_{nn}\right|^2 T_n^{\dagger}\right\rangle\right]=-\mathrm{Im}\left\langle T_n\bar{G}_{nn}^{\dagger}T_n^{\dagger}\right\rangle. \qquad (37)$$

Finally, using Eq. (32) we arrive at Eq. (34).

The identity (34) simplifies Eq. (33) which can now be written as follows

$$\gamma_n\frac{\mathrm{Im}\,\bar{G}_{nn}}{\mathrm{Im}\,\Sigma_n}=\sum_{l,m}\bar{G}_{nl}(\Sigma_R^I)_{lm}\bar{G}_{mn}^{\dagger}+\sum_{l}\bar{G}_{nl}\gamma_l\bar{G}_{ln}^{\dagger}. \qquad (38)$$

Now, we exploit explicitly the periodicity of our system in the plane of the layers, using a mixed representation $(\mathbf{k}_{\parallel},n)$, where $\mathbf{k}_{\parallel}$ is the transverse wave vector and $n$ is the layer number. Taking into account that $\Sigma_n$, $\Gamma_n$ and $T_n$ are independent of a site in the plane and assuming that $\langle n|\Sigma_R(\mathbf{k}_{\parallel})|m\rangle=\Sigma_R(\mathbf{k}_{\parallel})\delta_{nN}\delta_{mN}$, we rewrite Eq. (38) as follows

$$\gamma_n\frac{\mathrm{Im}\,\bar{G}_{nn}}{\mathrm{Im}\,\Sigma_n}=\int\bar{G}_{nN}(\mathbf{k}_{\parallel})\Sigma_R^I(\mathbf{k}_{\parallel})\bar{G}_{Nn}^{\dagger}(\mathbf{k}_{\parallel})\frac{d\mathbf{k}_{\parallel}}{(2\pi)^2}+\sum_m\gamma_m\int\bar{G}_{nm}(\mathbf{k}_{\parallel})\bar{G}_{mn}^{\dagger}(\mathbf{k}_{\parallel})\frac{d\mathbf{k}_{\parallel}}{(2\pi)^2}. \qquad (39)$$

Here $\bar{G}_{nn}$ is the on-site matrix element of the Green's function within layer $n$ given by Eq. (20). The transmission is given by

$$T=\int\Sigma_R^I(\mathbf{k}_{\parallel})\bar{G}_{1N}(\mathbf{k}_{\parallel})\Sigma_R^I(\mathbf{k}_{\parallel})\bar{G}_{N1}^{\dagger}(\mathbf{k}_{\parallel})\frac{d\mathbf{k}_{\parallel}}{(2\pi)^2}+\sum_m\gamma_m\int\Sigma_R^I(\mathbf{k}_{\parallel})\bar{G}_{1m}(\mathbf{k}_{\parallel})\bar{G}_{m1}^{+}(\mathbf{k}_{\parallel})\frac{d\mathbf{k}_{\parallel}}{(2\pi)^2}. \qquad (40)$$

Using definitions (6), expressions (39) and (40) can be rewritten as follows:

$$2\tilde{\gamma}_n\Sigma_n^I\,\mathrm{Im}\,\bar{G}_{nn}=T_{nR}+\sum_m\tilde{\gamma}_m T_{nm}, \qquad (41)$$

$$T=T_{LR}+\sum_m\tilde{\gamma}_m T_{Lm}, \qquad (42)$$

where

$$\gamma_n=\Sigma_n^I\tilde{\gamma}_n. \qquad (43)$$

Finally, we rewrite Eqs. (41) for the vertex functions $\tilde{\gamma}_n$ entirely in terms of transmission functions (6). For this purpose, we introduce the self-energy $\Sigma$ whose matrix elements are

$$\Sigma_{mn}(\mathbf{k}_{\parallel})=\Sigma_n(\mathbf{k}_{\parallel})\delta_{mn}=\Sigma_n\delta_{mn}+\Sigma_L(\mathbf{k}_{\parallel})\delta_{m1}\delta_{n1}+\Sigma_R(\mathbf{k}_{\parallel})\delta_{mN}\delta_{nN}. \qquad (44)$$

It is easy to see that

$$T_{nR}+T_{nL}+\sum_m T_{nm}=\Sigma_n^I\left[\bar{G}(2\,\mathrm{Im}\,\Sigma)\bar{G}^{\dagger}\right]_{nn}=2\Sigma_n^I\,\mathrm{Im}\,\bar{G}_{nn}, \qquad (45)$$

where the integration over $\mathbf{k}_\parallel$ is implicitly assumed in the operator product in the square brackets, and the latter equality follows from the identity: $\text{Im}\,\bar{G} = \bar{G}(\text{Im}\,\Sigma)\bar{G}^\dagger$. Using Eq. (45), we can rewrite Eqs.(41) as follows:

$$0 = \tilde{\gamma}_n T_{nL} + (\tilde{\gamma}_n - 1)T_{nR} + \sum_m (\tilde{\gamma}_n - \tilde{\gamma}_m)T_{nm}, \quad n = 1, 2, \ldots N. \tag{46}$$

These equations are identical to those given by formula (4) within the Büttiker's voltage-probe model. $\tilde{\gamma}_n$ can be associated with a relative chemical potential of a floating electrode $n$ measured with respect to the chemical potential of the left electrode set equal to zero, the chemical potential of the right electrode being set equal to unity. This simply implies that $\tilde{\gamma}_n$ is the reduced chemical potential given by $\tilde{\gamma}_n = (\mu_n - \mu_L)/eV$. The physical meaning of Eq. (42) is then the transmission to the left electrode from the right electrode and all the floating electrodes. The local current conservation condition requires that the local currents in all the floating electrodes are zero. This is exactly what Eqs. (46) infer. Thus, the CPA vertex constants for electric conductivity may be interpreted as local chemical potentials that provide zero current in the floating electrodes.

## 6. Examples and discussion

While within the voltage-probe model the self-energies of the floating electrodes are phenomenological parameters, the single-site CPA provides a clear recipe to determine the self-energies. Once the type of disorder is known the self-energies can be found according to a self-consistent procedure that provides on average a zero on-site *T*-matrix. In that sense, the proven equivalence between the CPA and the voltage-probe model may be considered as a concrete physical example where the voltage-probe model is justified.

The variation of the local chemical potential across the sample implies the presence of the internal electric field. It is known that such a field may be used instead of the vertex corrections for conductivity to provide the local current conservation.[36] For example, such an approach was used to calculate the conductance and magnetoresistance of segmented nanowires in the presence of diffuse scattering.[37] We have shown within the CPA how to calculate the local chemical potential and thus the internal electric field.

The local chemical potential allows obtaining a useful insight into the transport behavior within the CPA. The variation in the chemical potential across the conductor may be used to determine the local resistivity and thus identify regions in an inhomogeneous sample contributing differently to the resistance. This approach also allows finding the interface resistance between the regions separating two disordered conductors. Below we consider a few examples.

We calculate the conductance by considering a sample of disordered material which consists of $N$ layers and is connected to two perfect semi-infinite electrodes. We assume that the sample is a binary alloy characterized by on-site energy $U^A$ with a probability $q^A$ and on-site energy $U^B$ with a probability $q^B$. The coherent potential of the system can be found by solving self-consistently Eqs. (19). The Green's function of the sample connected to the electrodes within the CPA is given by Eq. (21). We assume a single-band tight-binding model and simple cubic lattice. In this case, the eigenvalues of the Hamiltonian $H$ are given by $\varepsilon(\mathbf{k}_\parallel) = -2t(\cos k_x a + \cos k_y a)$, where $t > 0$ is the hopping integral between neighboring sites and $a$ is the lattice constant. The self-energies of the electrodes are expressed through the surface Green's functions of the leads and are given by[38] $\Sigma_{L,R}(E, \mathbf{k}_{\parallel}) = \left\{ E \pm \sqrt{\left[ E - \varepsilon(\mathbf{k}_{\parallel}) \right]^2 - 4t^2} \right\} / 4$.

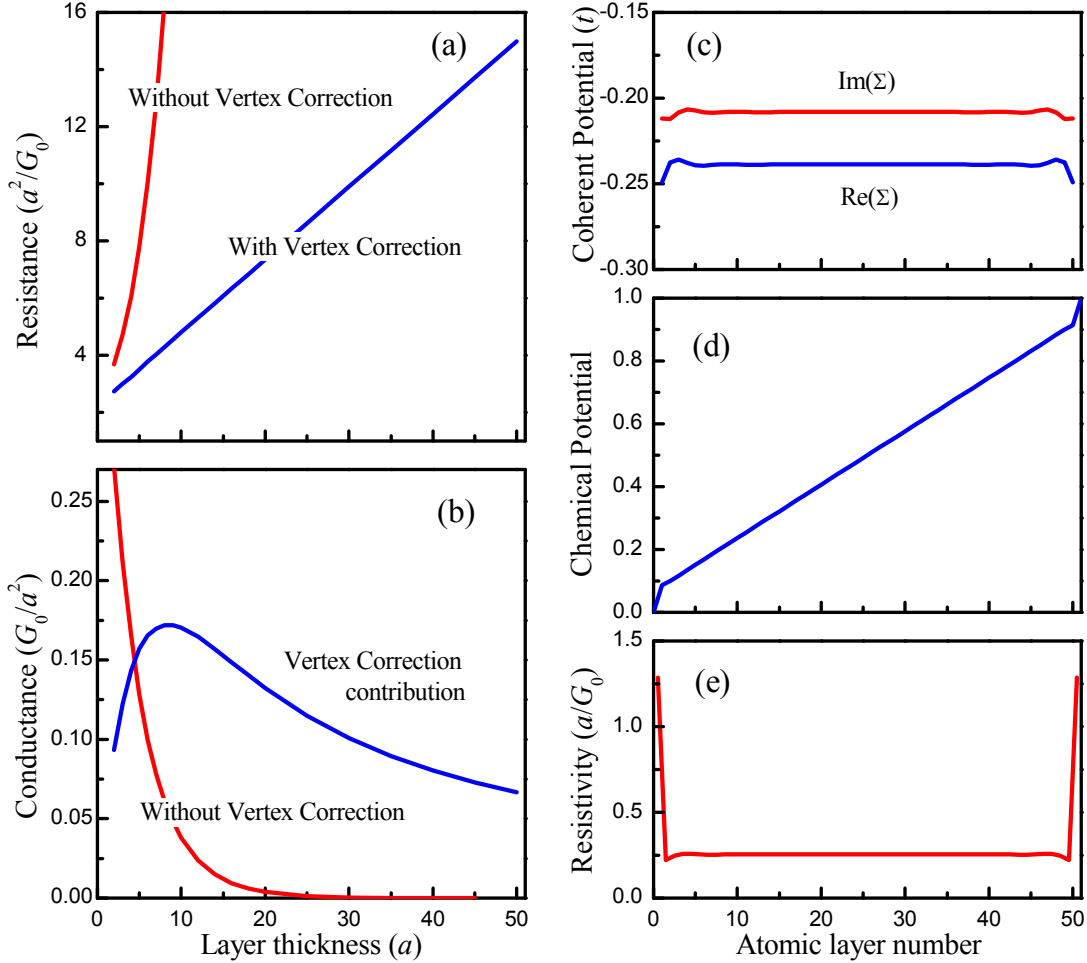

**Fig. 1:** Results of transport calculations for a binary alloy layer placed between two semi-infinite electrodes. (a) Areal resistance as a function of layer thickness with and without vertex corrections; (b) Areal conductance as a function of layer thickness showing separately the conductance without vertex corrections and the vertex correction contribution; (c) Real and imaginary parts of the coherent potential across the sample; (d) Reduced chemical potential and (e) local resistivity across the sample. Parameters used in the calculations: $E_F = -2t$, $U^A = -0.6t$, $U^B = 1.4t$, $q^A = 0.7$, $q^B = 0.3$. In figures (c), (d), and (e) the layer thickness is $50a$. $G_0 = 2e^2/h$ is the conductance quantum, and $a$ is the lattice constant.

Fig.1 shows results of the calculation for a binary alloy layer embedded between two semi-infinite electrodes. Here we set $E = E_F = -2t$, $U^A = -0.6t$, $U^B = 1.4t$, $q^A = 0.7$, $q^B = 0.3$ which provides a relatively weak disorder in the alloy. It is seen from Figs. 1a and 1b (red curves) that ignoring the vertex correction in the transport calculation leads to an exponential increase of the areal resistance $R$ (Fig. 1a) and decrease in the conductance $G$ per unit area (Fig. 1b) with disordered layer thickness. It has been shown that the CPA conductance without vertex corrections is similar to the ballistic contribution which conserves $\mathbf{k}_\parallel$ in the process of transmission across a disordered region.[39] This contribution decreases exponentially on a scale determined by the mean free path as determined by imaginary part of the CPA self-energy (the blue curve in Fig. 1c). The vertex corrections restore the Ohm's law making the resistance to increase linear with layer thickness (the blue curve in Fig. 1a). As is seen from Fig. 1b, the vertex contribution first increases with layer thickness, reaches maximum, and then decreases inversely proportional to layer thickness. This behavior reflects a diffusive contribution to the conductance, where elastic scattering involves scattering events between different $\mathbf{k}_\parallel$ resulting in the increase of the diffusive part on the scale of the mean free part. Further increasing of the layer thickness enhances the diffusive contribution to the resistance proportional to the number of scattering events (layer thickness).

Since the alloy is assumed to be homogeneous the only inhomogeneity in the system occurs near the interfaces between the disordered layer and perfect electrodes. This is reflected in the coherent potential which is nearly constant across the layer, small variations being seen only near the interfaces (Fig. 1c). The homogeneity of the bulk alloy is mirrored in the chemical potential variation which drops linearly across the disordered region (Fig. 1d). The only sizable deviation from the linear behavior occurs at the interfaces with electrodes where steps in the chemical potential reflect the interface resistance (layers 0-1 and 50-51 in Fig. 1d).

In general, the chemical potential profile across a disordered inhomogeneous conductor may be used to evaluate the local resistivity of the conductor, which may be useful for analyzing the transport behavior. Since our system is quasi one-dimensional and the current is conserved, we can define the local resistivity as follows:

$$\rho(z) = \frac{d\tilde{\gamma}}{dz} R, \tag{47}$$

where $R$ is the areal resistance of the whole system and it is assumed that the reduced chemical potential is the continuous function of position $z$ across the conductor. In our case of a discrete lattice we can define the local resistivity, e.g., as follows $\rho_n = (\tilde{\gamma}_{n+1} - 2\tilde{\gamma}_n + \tilde{\gamma}_{n-1}) R/a$. Fig. 1e shows, as an example, the result of calculation of the resistivity for the system discussed above. We see that the resistivity is nearly constant through the disordered layer, but has sharp features near interfaces reflecting the interface resistance. There are weak oscillations in the resistivity near the interfaces reflecting the quantum interference caused by interface perturbation.

In another example we consider a diffusive bilayer conductor representing two disordered alloy layers (25 unit cells each) placed between two semi-infinite electrodes. Here we assume that $E_F = -2t$ and disorder in the left ($L$) layer is fixed so that $U_L^A = -3t$, $U_L^B = t$, $q_L^A = 0.35$, and $q_L^B = 0.65$. The on-site atomic energies in the right ($R$) disordered layer are also fixed, $U_R^A = 3t$ and $U_R^B = -t$, whereas the relative concentration of the two alloy components, $q_R^A$ and $q_R^B$, is varied. The results for the reduced chemical potential are displayed in Fig. 2a for different values of $q_R^A$. It is seen that with increasing disorder the voltage drop in the right segment becomes more pronounced reflecting the increasing resistivity of this layer. This is evident from Fig. 2b showing the site-dependent resistivity across the conductor. The resistivity of the left layer $\rho_L$ is constant, whereas the resistivity of the right layer $\rho_R$ increases with alloying.

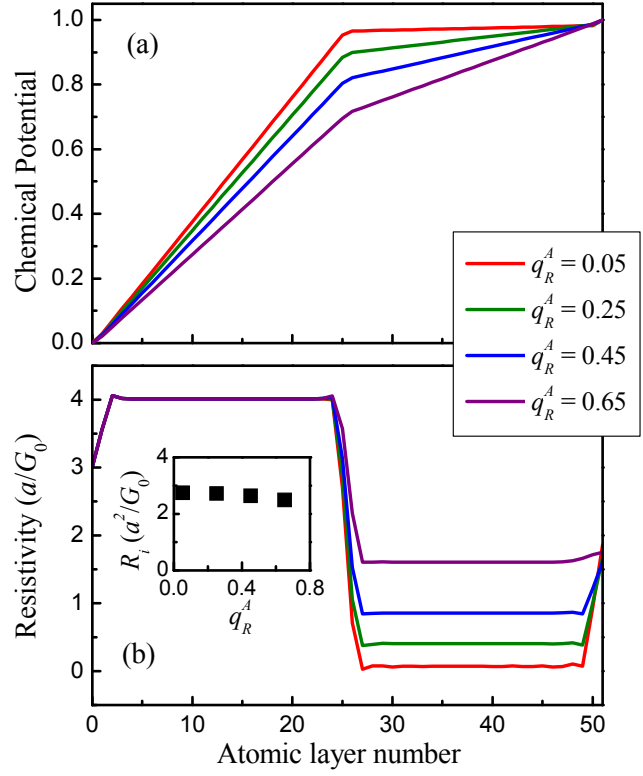

**Fig. 2:** Results of transport calculations for a disordered bilayer system. (a) Reduced chemical potential across the bilayer conductor for different concentration $q_R^A$ in the right segment. (b) Local resistivity across the bilayer. The inset shows the interface resistance between two disordered layers as a function of alloying in the right segment. Parameters used in the calculations: $E_F = -2t$, $U_L^A = -3t$, $U_L^B = t$, $q_L^A = 0.35$, $q_L^B = 0.65$, $U_R^A = 3t$, and $U_R^B = -t$. $G_0 = 2e^2/h$ is the conductance quantum, and $a$ is the lattice constant.

These results allow us to evaluate the interface resistance between two disordered layers. The interface resistance has previously been derived in terms of transmission probabilities between two ballistic electrodes assuming completely diffuse scattering in the bulk of the layers.[40] In our approach the diffuse scattering in the two adjacent layers is provided by the CPA. In order to calculate the interface resistance, we fix two points in the conductor lying at distance $t_L$ from the interface in the left layer and at distance $t_R$ from the interface in the right layer. The areal resistance of the sample between the two points can be written as follows:

$$R \Delta \tilde{\gamma} = \rho_L t_L + R_i + \rho_R t_R, \tag{48}$$

where $R$ is the areal resistance of the whole system and $R_i$ is the interface resistance. By fixing $t_L$ and $t_R$, so that the two points lie sufficiently far away from the interface and hence the local resistivities of the left ($\rho_L$) and right ($\rho_R$) layers are nearly the same as in the bulk of these layers, we can calculate the interface resistance from Eq. (48) given the known value of $\Delta\tilde{\gamma}$ between the two points. The result is displayed in the inset of Fig. 2b which shows $R_i$ as a function of the degree of alloying in the right layer. It is known that depending on reflection coefficients diffuse scattering can assist or suppress conduction across interfaces.[41] In our case the interface resistance slightly decreases with disorder which is due to opening new transmission channels across the interface.[42] This approach to calculate the interface resistance in the presence of disorder may be considered as an alternative to that based on the supercell calculation[43,44] and the Boltzmann equation.[45]

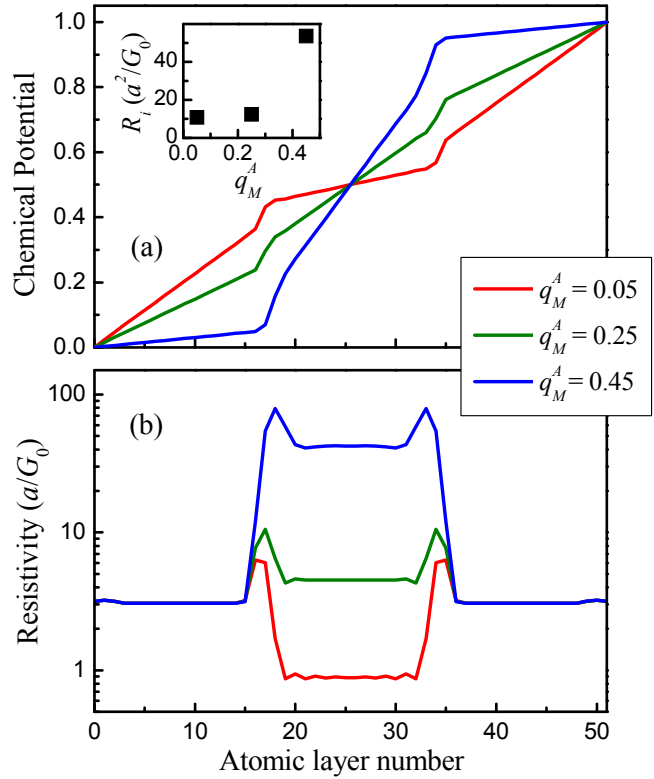

**Fig. 3:** Results of transport calculations for a disordered trilayer system. (a) Reduced chemical potential across the trilayer conductor system for different concentration $q_M^A$ in the middle segment. (b) Local resistivity across the trilayer. The inset shows the interface resistance between the left (right) and middle disordered layers as a function of alloying in the middle layer. The parameters used in the calculation are as follows: $U_{L,R}^A = -3t$, $U_{L,R}^B = t$, $q_{L,R}^A = 0.6$, $q_{L,R}^B = 0.6$, $U_M^A = -4t$, and $U_M^B = 2t$. $G_0 = 2e^2/h$ is the conductance quantum, and $a$ is the lattice constant.

Finally, we consider a disordered trilayer system where the left (*L*) and right (*R*) conducting layers of thickness 16*a* are separated by a conducting middle (*M*) layer of thickness 18*a*. The parameters characterizing the left and right layers are assumed to be identical, i.e. $U_{L,R}^A = -3t$, $U_{L,R}^B = t$, $q_{L,R}^A = 0.4$, and $q_{L,R}^B = 0.6$. We study transport properties of the system as a function of alloying $q_M^A$ in the middle layer for which the on-site atomic energies of the alloy components are assumed to be $U_M^A = -4t$ and $U_M^B = 2t$. Fig. 3a shows the resulting variation of the reduced chemical potential across the trilayer. With increasing $q_M^A$ the voltage drop across the middle layer is increasing reflecting the increasing resistivity of this layer. The latter fact is also evident from the site-dependent resistivity plots shown in Fig. 3b. Steps are seen at the interfaces between the middle and adjacent layers as the result of the interface resistance. By performing a calculation similar to that for the bilayer system we find that in this case the interface resistance increases significantly with concentration $q_M^A$ in the middle layer alloy (see the inset in Fig. 3a). This is due to a large mismatch between the on-site energy $U_M^A$ and the respective on-site energies in the left (right) layer alloys which leads to the large potential step at the interface with increasing $q_M^A$.

## 7. Summary

This work links the coherent potential approximation that has been widely used to describe the residual resistivity of binary alloys to the Büttiker's voltage-probe model that has been developed to treat phase breaking scattering in mesoscopic systems. The CPA is typically applied to treat the conductivity due to elastic scattering originating from substitutional disorder. For a given configuration, the electronic transport remains coherent. Configurational averaging replaces the original phase-coherent system by the one involving energy level broadening similar to that occurring as a result of the coupling to reservoirs that breaks coherence in electron transmission and produces inelastic scattering. In that sense, the CPA has an analogy to the Büttiker's voltage-probe model, though the latter was introduced for a different purpose, namely to take into account in a simple way phase breaking scattering in mesoscopic conductors that is essential in experimental conditions.

Within both methods just adding on-site self-energies within the Landauer-Büttiker approach for conductance would lead to current dissipation. To provide the local current conservation the chemical potentials of the voltage probes need to be adjusted to guarantee no current in the floating electrodes. We have shown that within the CPA this procedure is equivalent to taking into account the vertex corrections, and the spatial dependence of the vertex exactly follows the local chemical potential. This interpretation allows the determination of the chemical potential profile across a disordered conductor which is useful for analyzing results of transport calculations within the CPA. In particular, for layered systems with translational periodicity in the plane of the layers one can introduce the local resistivity which reflects the distribution of the resistance across the conductor. The method also allows calculating the interface resistance between disordered layers. This approach has been illustrated by considering examples of single-layer, bilayer, and trilayer conductors consisting of different disordered binary alloys within a tight-binding model. The proposed method may be extended to multiband spin-dependent systems[46] and applied to real geometries of disorder multilayers.[47]

**Acknowledgements**

M.Ye.Zh. thanks the Department of Physics and Astronomy at the University of Nebraska-Lincoln for hospitality during his stay in spring 2011. E.Y.T. thanks Gerrit Bauer for discussing the results of this work. This research was supported by the National Science Foundation through the Nebraska EPSCoR Track II project (Grant No. EPS-1010674) and Nebraska MRSEC (Grant No. DMR-0820521).

* E-mail: tsymbal@unl.edu